\documentclass[useAMS,usenatbib]{mn2cls/mn2e}
\usepackage{epsfig,amsmath,natbib}

\def\gtsima{$\; \buildrel > \over \sim \;$}
\def\ltsima{$\; \buildrel < \over \sim \;$}
\def\gsim{\lower.5ex\hbox{\gtsima}}
\def\lsim{\lower.5ex\hbox{\ltsima}}

\def\lcdm{{$\Lambda$CDM }}

\def\de{{\mathrm{d}}}

\def\Spec{{\mathcal{E}}}
\def\Source{{\mathcal{S}}}

\title[Turbulence in the IGM]{Turbulence in the Intergalactic Medium}
\author[Evoli \& Ferrara]
{Carmelo Evoli$^{1}$\thanks{E-mail: evoli@sissa.it} \& Andrea Ferrara$^{2}$ \\
$^{1}$ SISSA, Via Bonomea 265, 34136 Trieste, Italy \\
$^{2}$ Scuola Normale Superiore, Piazza dei Cavalieri 7, 56127, Pisa, Italy}
\date{}
\pagerange{\pageref{firstpage}--\pageref{lastpage}} 
\pubyear{2011}

\begin{document}

\maketitle
\label{firstpage}

\small

\begin{abstract}
We study supernova-driven galactic outflows as a mechanism for injecting turbulence in the intergalactic medium (IGM) far from galaxies. To this aim we follow the evolution of a $10^{13}$~M$_\odot$ galaxy along its merger tree, with carefully calibrated prescriptions for star formation and wind efficiencies. At $z\approx 3$ the majority of the bubbles around galaxies are old (ages $> 1$~Gyr), i.e. they contain metals expelled by their progenitors at earlier times; their filling factor increases with time reaching about 10\% at $z < 2$. The energy deposited by these expanding shocks in the IGM is predominantly in kinetic form (mean energy density of 1 $\mu$eV cm$^{-3}$, about $2-3 \times$ the thermal one), which  is rapidly converted in disordered motions by instabilities, finally resulting in a fully developed turbulent spectrum whose evolution is followed through a spectral transfer function approach. The derived {\it mean} IGM turbulent Doppler parameter, $b_t$, peaks at $z\approx 1$ at about 1.5~km~s$^{-1}$ with maximum $b_t = 25$~km~s$^{-1}$. The shape of the $b_t$ distribution does not significantly evolve with redshift but undergoes a continuous shift towards lower $b_t$ values with time, as a result of bubble aging. We find also a clear trend of decreasing $b_t$ with $N_{\rm HI}$ and a more complex dependence on $R_s$ resulting from the age spread of the bubbles. We have attempted a preliminary comparison with the data, hampered by the scarcity of the latter and by the challenge provided by the subtraction of peculiar and thermal motions. Finally we comment on the implications of turbulence for various cosmological studies. 
\end{abstract}



\section[]{Introduction}

The intergalactic medium (IGM) is a pervasive, diffuse cosmic baryonic component out of which galaxies form by accretion. In turn, the IGM is replenished with gas and heavy elements carried by galactic outflows. Such dynamical processes hence encode a record of the complex galaxy-IGM interplay. 

Observations of metal absorption lines (e.g. C~III , C~IV, Si~III, S~IV and O~VI) in quasar spectra are a direct probe of this evolution. They show that regions of enhanced IGM density far from large galaxies are polluted with non-negligible amounts of heavy elements \citep[e.g.][]{Songaila96, Dave98, Schaye00a, Pettini03, Schaye03, Aguirre04, Aracil04, Simcoe04}. In addition, Ly$\alpha$ forest clouds with neutral hydrogen column density $\log N(H I) > 14$ are metal enriched to $Z \approx 10^{-3}-10^{-2}$ Z$_\odot$ \citep{Cowie95, Songaila96, Ellison00}; finally, the carbon and silicon abundances of such systems remain roughly constant throughout the redshift range $1.5 < z < 5.5$ \citep{Songaila01, Pettini03}, although the first signs of a slight decrease above $z=6$ have been now tentatively reported \citep{Ryan-Weber09}. More recently, at variance with the constant behaviour picture, \cite{Dodorico10} reported a significant increase of the C~IV abundance towards lower redshifts $z \lsim 2$. This result has been also confirmed by \citet{Cooksey10} for $z \lsim 1$.

Since metals are created in stars inside galaxies, their transport into these distant regions must rely on some yet unknown mechanisms. The most likely one is supernova-driven galactic outflows (\citealp{MacLow99}; \citealp{FPS} [hereafter FPS]; \citealp{Aguirre01, Oppenheimer06, Cen10}), although different alternatives have been proposed, i.e. dust-sputtering \citep{Aguirre01b,Bianchi05} or merging \citep{Gnedin98}. This hypothesis is supported by a series of observations of galactic winds \citep[e.g.][]{Heckman01,Pettini01,Adelberger03,Shapley03,Erb06} from which one can conclude that (on some spatial scale) galactic wind velocities range from hundreds to thousands of km/s and the mass loss rates are comparable to star formation rates (SFRs).

A complementary piece of information comes from the well-established fact that the interstellar medium (ISM) in galactic disks is turbulent \citep{Larson81,Scalo87,Dickey90,Elmegreen04}. In the Milky Way (MW), the observed vertical distribution of gas, and especially the cold neutral HI component, cannot be supported by thermal pressure alone \citep{Lockman91,Ferrara93} and an additional pressure contribution must be advocated. Furthermore, pulsar scintillation experiments have convincingly demonstrated that electron density fluctuations in the  ionized ISM are characterized by a Kolmogorov spectrum on scales ranging from AUs to kpc \citep{Armstrong95, Han04}. 
The Kolmogorov spectrum is the spectrum of the density fluctuactions which arise from turbulent motions in a compressible gas with constant dynamical energy exchange between irregularities on different scales and, under very general conditions, it assumes the form of a power-law ($\sim k^{-11/3}$ where $k$ is the wave-number associated to a given scale $L$) within the inertial range. 
In other spiral galaxies the vertical velocity dispersions (derived essentially from HI observations) are observed to vary radially from $12-15$ km/s in the central parts to $4-6$~km/s in the outer regions and, in most cases, these values exceed those expected by the thermal broadening of the HI emission line \citep{Dib06}. 

Often invoked to provide the required additional pressure are turbulent motions of the interstellar gas, magnetic fields, and cosmic rays \citep{Boulares90,Lockman91,Dib06,Joung09}. Turbulence is likely to be to most relevant pressure contribution; its driving may be due to different sources (star formation, stellar outflows, instabilities) acting on specific characteristic scales and injecting different amounts of kinetic energy into the medium;  among these, supernova (SN) explosions are thought to be the dominant agent, at least in the star forming regions of galaxies \citep{McKee77,Norman96,MacLow04,deAvillez07}. If both turbulence and metal enrichment are predominantly driven by the same physical phenomenon, i.e. kinetic energy deposition by supernova shocks, then a natural expectation is that IGM gas in enriched regions should show signatures of a turbulent regime. 

A direct way to measure turbulence in the IGM is to look for velocity differences on the smallest spatial scales accessible to observations; to this scope, \citet{Rauch01} in a pioneering experiment used lensed quasars. They observed adjacent CIV profiles in paired lines of sights, and find velocity differences of $\delta v \approx 5$~km~s$^{-1}$ on scales of $300$~pc at redshift $\sim 2.7$. They apply the Kolmogorov steady-state assumption to calculate an energy transfer rate $\epsilon \approx 10^{-3}$~cm$^2$~sec$^{-3}$ and a turbulent dissipation timescale as short as $\approx 10^8$~yrs, thus implying a recurrent stirring of the gas by some sort of galactic-scale feedback. 

Another hint of a significant turbulent contribution to the IGM kinetic budget comes from the fact that the median Doppler parameters measured in the Ly$\alpha$ forest are significantly larger than those predicted by cosmological simulations. Again, this implies that some energy in non-thermal form must be injected in the gas to explain the observed line broadening \citep{Meiksin01, Meiksin09}.

Quite surprisingly, the properties of intergalactic turbulence have received so far relatively little attention in spite of the insights that its detailed understanding might provide on the galaxy-IGM interplay (FPS; \citealp{Oppenheimer09}; \citealp{Scannapieco10}). \citet{Oppenheimer09} trying to model low-$z$ OVI absorbers concluded that their properties fit observables if and only if sub-resolution turbulence (in practice a density dependent turbulent $b$-parameter) is added at a level which increases with OVI absorber strength (and hence presumably closer to high density regions hosting galaxies). In addition, stronger absorbers arise from more recent outflows. It is important to note, that turbulence, if present, can also profoundly modify the enrichment patterns through diffusion processes, as pointed out already by FPS. Additional important information might come from investigations of structure formation, during which the released gravitational energy is transferred to the IGM in different forms (e.g. gas entropy, instabilities, cosmic ray acceleration); in particular,  turbulence is induced by the vorticity cascade originating at cosmological shocks. This has been investigated throughout cosmological simulations by \citet{Ryu08} and \citet{Zhu10} respectively, who derived the average magnetic field strength and turbulent pressure in the overdense IGM regions outside clusters/groups.

In this work, we try to fill the aforementioned gap by investigating in detail the supernova-driven outflow scenario as a mechanism to pump and sustain turbulence in the IGM surrounding high-redshift galaxies. We investigate the evolution of winds via a semi-analytic approach following both galaxy evolution along its hierarchical growth and the expansion of supernova driven superbubbles as they escape from the halo potential well. With such results in hand, we model turbulence evolution within wind shells using a novel approach based on the spectral transfer equation. This approach allows to derive the turbulent energy density deposited by galactic winds in the IGM, along with its spectral features and dissipation time scales, and to predict the corresponding thermal/kinetic evolution of the IGM.


\section[]{THE MODEL}

In this Section we describe how we determine the evolution of the turbulent energy of the deposited by galactic outflows into the IGM. First, we introduce the galaxy formation model, allowing a precise description of both the SF history and the SN feedback along the galaxy merger tree (MT). These results are then used as inputs to model the evolution of pressure-supported bubbles through the solution of dynamical equations and the turbulence energy spectrum evolution within them.   

\subsection{Cosmic star formation history}

The semi-analytic model we introduce in this Section to describe the evolution of a galaxy along its MT is similar to the one-zone model by \citet{Salvadori07} with the improvements introduced by \citet{Salvadori08}. Since in this work we are not interested in following the chemical enrichment in detail, we use a simplified approach described in the following. 

In hierarchical models of structure formation, such as $\Lambda$CDM, the formation of a DM halo through accretion and repeated mergers can be described by a MT \citep{Lacey93}.
MTs, which list the progenitors of a given halo at different redshifts and describing how and when these merge together, contain essentially all the necessary information about the dark matter content of halos to build realistic model of galaxy formation as demonstrated by a long-standing practice in the field \citep[e.g.][]{Kauffmann93,Somerville99}.

The MTs can be extracted from N-body simulations \citep{Springel05} or generated by Monte-Carlo algorithms \citep[e.g]{Sheth99}. In this work we compute the mass growth histories of DM halos using the second method using a public code\footnote{\tt http://star-www.dur.ac.uk/\textasciitilde cole/merger\_trees} based on the  extended Press-Schecter formalism \citep{Cole00}, as improved by \citet{Parkinson08}, to which we defer the reader in numerical details. As an input, we use a WMAP5\footnote{Throughout all the paper we adopt a flat $\Lambda$CDM cosmological model with $h=0.72$, $\Omega_{b}=0.044$, $\Omega_{m}=0.26$, $n_{s}=0.96$ and $\sigma_{8}=0.8$ consistent with the five-year WMAP parameter analysis \citep{WMAP5}.} cosmological parameters and power-spectrum. As an output, we obtain many realizations of the MT for a given halo mass $M_0$ at a certain final redshift $z_{\rm f}$; each realization lists all progenitors of $M_0$ at different redshifts, following the related merging histories down to a mass of $M_{\rm res}$.

In our model all the proto-galaxies virializes with a gas-to-dark matter ratio of 1:5. The neutral gas in these mini-halos cannot cool via atomic hydrogen and relies on the presence of molecular hydrogen, H$_{2}$, to cool and collapse, ultimately forming stars. Since we assume that feedback effects rapidly suppress star formation in the first minihalos and that only Ly$\alpha$ cooling halos ($T_{\rm vir} > 10^4 K$) contribute to the SF history of the galaxy, we choose $M_{\rm res} = M_4(z=20)$ where $M_4(z)$ is the mass corresponding to a halo with virial temperature of $10^4$~K at redshift $z$, given 
\begin{equation}
M_4(z) \approx 10^8 M_{\odot} \left( \frac{10}{1+z} \right)^{3/2} 
\end{equation}
(for an exact expression see \citet{Barkana01}) and $z=20$ is the starting redshift of our simulation. Note that the value of $M_{\rm res}$ sets the limit between \emph{progenitors} and \emph{mass accretion}, since the galaxies above $M_{\rm res}$ experience star formation and stellar feedback, ultimately changing their gas content, whereas objects below this threshold retain their original gas content fraction during their evolution, which is finally inherited by another halo in the next hierarchy level. 

Star formation in gas clouds occurs on a free-fall timescale $t_{\rm ff} = (3 \pi/32 G \rho)^{1/2}$ where $G$ is the gravitational constant and $\rho$ is the average mass  (dark+baryonic) density inside the halo assumed to have a virial radius ($r_{\rm vir}$) as given in \citet{Bryan98}. The star formation efficiency of a galaxy is then modeled as a fraction, $\epsilon_*$, of the free-fall time:
\begin{equation}\label{sfr}
\psi (z) = \epsilon_* \frac{M_{\rm g}(z)}{t_{\rm ff}(z)}
\end{equation}
where $\epsilon_*$, represents a free parameter of the model and $M_{\rm g}$ represents the gas mass in the halo which has not yet been converted into stars.

According to the standard scenario for galaxy formation the gas inside the galaxy is depleted by various feedback processes, the most important being gas-loss driven by SN explosions. In fact, SN explosions may power a wind which, if sufficiently energetic, may overcome the gravitational pull of the host halo leading to expulsion of gas and metals into the surrounding IGM. To model this process, we compare the kinetic energy injected by SN-driven winds with the minimum kinetic energy of a mass $M_{\rm w}$ to \emph{escape} the galactic potential well \citep{Larson74,Dekel86,Ferrara00}. The mass of gas ejected from the galaxy is computed from the equation:
\begin{equation}\label{eje}
\frac{1}{2} M_{\rm w} v_e^2 = E_{SN}
\end{equation}
where
\begin{equation}\label{ESN}
E_{SN} =  \epsilon_{\rm w} N_{SN} \langle E_{SN} \rangle
\end{equation}
is the kinetic energy injected by SNe and $v_e^2 = 2 E_b/M$ is the escape velocity of the gas from a halo with mass $M_h$ and binding energy $E_b$ given by \citet{Barkana01}:
\begin{equation}
E_b = \frac{1}{2} \frac{GM^2}{R_{vir}} = 5.45 \times 10^{53} \left( \frac{M_h}{10^8 h^{-1} M_\odot} \right)^{5/3} \left( \frac{1+z}{10} \right) h^{-1} \, {\rm erg} \,.
\end{equation}

In Eq.~\ref{ESN}, $\epsilon_{\rm w}$ is a free parameter which controls the conversion efficiency of SN explosion energy in kinetic form, $N_{SN}$ is the number of SN, and $\langle E_{SN} \rangle = 1.2 \times 10^{51}$ erg is the average SN explosion energy (not in neutrinos). Differentiating with respect to time Eq.~\ref{eje} we find that the gas ejection rate is proportional to the SN explosion rate:
\begin{equation}\label{mej}
\dot{M}_{\rm w} = \frac{2 \epsilon_{\rm w} \langle E_{SN} \rangle}{v_e^2} \dot{N}_{SN}
\end{equation}
In our model we only consider Type II SNe (pair-instability supernovae are neglected as  PopIII stars do not contribute significantly to the total SFR \citep{Tornatore07}); hence we calculate $N_{\rm SN}$ by integrating the adopted IMF (see below) in the canonical supernova range 8-100 $M_\odot$.  For any star forming halo of the galaxy hierarchy, we therefore solve the following system of differential equations:
\begin{eqnarray}
\dot{M_*}(z) & = & \left[ 1-R(z) \right] \, \psi(z) \\
\dot{M_{\rm g}}(z) & = & \dot{M}_a(z) - \left[ 1-R(z) \right] \, \psi(z) - \dot{M}_{\rm w} (z)
\end{eqnarray}
The first equation simply defines the SFR. In our model, we have assumed the Instantaneous Recycling Approximation \citep[IRA][]{Tinsley80}, according to which stars are divided in two classes: those which live forever, if their lifetime is longer than the time since their formation and those which die instantaneously, eventually leaving a remnant. The transition mass between the two possible evolutions, or turn-off mass ($m_{\rm to}$), has been computed at any considered redshift. Under IRA, the \emph{returned fraction}, that is the stellar mass fraction returned to the gas through winds and SN explosions is:
\begin{equation}
R(z) = \frac{\int_{m_{\rm to}(z)}^{100 M_\odot} [m-w_m(m)] \phi(m) dm} {\int_{0.1 M_\odot}^{100 M_\odot} m \phi(m) dm}
\end{equation}
where $\phi(m)$ is the IMF of the newborn PopII/I stars and it is assumed to have the form:
\begin{equation}
\phi(m) = \frac{\de N}{\de m} \propto m^{-1+x} \exp \left( -\frac{m_{\rm cut}}{m} \right)
\end{equation}
with $x=-1.35$, $m_{\rm cut} = 0.35$~M$_{\odot}$ and $m$ in the range $[0.1,100]$~M$_\odot$ \citep{Larson98}. The quantity $w_{m}(m)$ represents the mass of the stellar remnant left by a star of mass $m$ which explodes as SN. We have used the grid of models by \citep{vandenHoek97} for intermediate mass stars ($0.9<m<8$~M$_{\odot}$) and \citep{Woosley95} for massive stars ($8<m<40$~M$_{\odot}$). With the adopted IMF the corrispondent SN energy per unit stellar mass formed is $1.36 \times 10^{49}$~ergs/M$_{\odot}$.

The second equation (Eq.~8) describes the mass variation of cold gas: the latter increases due to \emph{accretion} ($\dot{M}_a(z)$) and it decreases both because of astration and mass loss due to SN winds. To model gas accretion we assume that if a new halo virializes out of the IGM gas it forms with a cosmological gas fraction $f_{b}=\Omega_{b}/\Omega_{m} = 1/6$, whereas if it results form the merging of two already existing halos its gas content is simply the sum of the gas mass of the progenitors. Similarly, during the merger, the stellar mass of the new galaxy is assumed to be the sum of the stellar masses of the progenitors. The model free parameters ($\epsilon_{\rm w}$ and $\epsilon_*$) are fixed to match the global properties of the MW as we will discuss in more details in the next Section.

\subsection{Bubble evolution}\label{Sub:bubble}

In our simulation galactic outflows are treated as pressure-driven bubbles of hot gas emerging from star-forming galaxies. They expand working against IGM pressure, and are driven by the energy injected by multi-SN explosions. Most of the swept-up mass, both in the early adiabatic and in the following radiative phases, is concentrated in a dense shell bounding the hot over-pressurized interior.

For this reason, galactic bubbles are canonically studied by using the thin-shell approximation in which the shell expansion is driven by the internal energy ($E_{b}$) of the hot bubble gas \citep{Ostriker88, Madau01, Bertone05, Samui08}.

In this work we follow the modelization already presented in \cite{Madau01} (see their Sec.~2). However, in our model we derive the SFR evolution from a constrained galactic model and from the SFR we derive the mechanical luminosity, defined as $L(t)=\de E/\de t$ where $E$ is the energy produced by the total contribution of $N_{SN}$ SNe with energy $\langle E_{SN} \rangle$ and an efficiency $\epsilon_{\rm w}$.

In the following, we remind the reader of the relevant equations used to follow the evolution of the interesting bubble properties, namely the bubble radius $R_{s}$, the shock velocity $V_{s}$ and the internal bubble temperature $T_{b}$:
\begin{gather}
\label{dvdt} \frac{d}{dt} (V_s \rho \dot{R}_s)= 4 \pi R_s^2 (P_b - P) - \frac{GM(R_s)}{R_s^2} \rho V_s \\
\label{dedt} \frac{dE_b}{dt} = L(t) - 4 \pi R_s^2 P_b \dot{R}_s - V_s \bar{n}^2_{H,b} \Lambda(\bar{T}_b) \\
\label{dTdt} \frac{\de T_{b}}{\de t} = 3 \frac{T_{b}}{R_{s}} \dot{R_{s}} + \frac{T_{b}}{P_{b}} \dot{P_{b}} - \frac{23}{10} \frac{C_{1}}{C_{2}} \frac{kT_{b}^{9/2}}{R_{s}^{2}P_{b}}
\end{gather}
where $G$ is the gravitational constant; $M(R)$ is the galactic DM mass assumed to follow a NFW profile within a radius $R$; $\rho$ ($P$) is the density (pressure) of the ambient medium; $V_{s} = (4\pi/3) R_{s}^{3}$ is the volume enclosed by the shell; $P_b = E_b/2\pi R_s^3$ \footnote{This result has been derived by substituting the expression for the internal energy $E_{\rm th} = 3/2 N k_{b}T$ into the ideal gas law $P=nRT/V$.} is the pressure of the bubble gas assumed to be isothermal and with adiabatic index $5/3$; $\bar{n}_{H,b}^{2} \Lambda(\bar{T})$ is the cooling rate per unit volume of the hot bubble gas (whose average hydrogen density and temperature are  $\bar{n}_{H,b}$ and $\bar{T}$, respectively); $C_{1}=16\pi \mu m_{p} \eta / 25 k$ and $C_{2}= (125/39) \pi \mu m_{p}$ (where $\eta= 6 \times 10^{-7}$ (c.g.s. units) is the classical Spitzer thermal conduction coefficient, assuming a Coulomb logarithm equal to $30$, and $\mu = 0.59$ is the mean molecular weight of a primordial ionized H/He mixture). 

Finally, the cooling function (represented by $\Lambda(T)$ in Eq.~\ref{dedt}) depends on the hot bubble density, temperature and metallicity ($Z$). In principle, one could compute an average metallicity of the bubble, but for simplicity we assume for this quantity a constant value $Z= 0.1 Z_\odot$, noting that the final results are very slightly dependent on this choice within a reasonable range. The cooling rates due to gas radiative processes  are taken from \citet{Sutherland93}; the other relevant cooling agent is inverse Compton cooling off CMB photons \citep{Ikeuchi86}, which is dominant at higher redshifts.

The first two equations (Eq.~\ref{dvdt} and \ref{dedt}) derive from momentum and energy conservation. In particular, the right-hand side of Eq.~\ref{dvdt} represents the momentum gained by the shell from the SN-shocked wind which expands against external pressure and gravitational attraction; the right-hand side of Eq.~\ref{dedt} describes the mechanical energy input, the work done against the shell, and the energy losses due to radiation. The third equation (Eq.~\ref{dTdt}) is obtained by equating the rate at which the gas is injected from the shell into the cavity with the conductive evaporation rate. Note that, as the bubbles sizes are always much smaller that the horizon scale, the cosmological terms in the above equations can be safely neglected. 

Another important quantity to determine the fate of the wind, is the total mass in the shell, $M_{s}$, since the energy required to accelerate it increases with its mass. The shell grow rate $\dot M_s$ is wind mass deposition rate (Eq.~\ref{eje}) and the rate at which gas is  swept-up from the galactic environment (for $R_s \le r_{\rm vir}$) or from the IGM ($R_s>R_{vir}$):
\begin{equation}
\frac{\de M_s}{\de t} = \dot{M}_w + 4 \pi \rho R_s^2 \frac{\de R_s}{\de t}
\end{equation}

The ambient gas density is taken to be equal to the halo gas density distribution (assumed to have an isothermal profile) within $R_{vir}$, and to the IGM background density outside the virial radius. To determine the pressure of the ambient medium we must further assume the ambient gas behave as an ideal isothermal gas which, inside halos, is at the virial temperature \cite{Navarro96} and outside $R_{\rm vir}$, the IGM, is taken to be photo-heated by the UV-background radiation to temperature calculated as in e.g. \citet{Choudhury06}. Note that, even prior to reionization, the IGM into which the bubble expand will be locally ionized and heated to roughly the same temperature by the SN progenitor stars. 

However, we notice here that the bubble reaches very rapidly the virial radius, so that these bubbles are practically expanding in the IGM for most of their time. To see this, we consider as in \citet{Madau01} a typical objects virializing at high-redshift ($z=9$) with an halo mass  $M=10^{8} h^{-1}$~M$_{\odot}$. At these epochs, the dark matter halo of a subgalactic system will be characterized by a virial radius
\begin{equation}
R_{\rm vir} = 0.76 M_{8}^{1/3} h^{-1} \left( \frac{1+z}{10} \right)^{-1} \, \, \, {\rm kpc}
\end{equation}
The Sedov solution (which is a good approximation of the equation systems when the ambient gas pressure, gravity and cooling can be neglected) predicts that the shell radius evolves according to: $R_{s}= (125/154\pi)^{1/5} (Lt^{3}/\rho)^{1/5}$; assuming a constant luminosity of $10^{38}$~erg$^{-1}$, we find that the time taken to reach $R_{\rm vir}$ is $\sim 10^{7}$~yr. This time happens to be smaller than the Hubble time at that redshift.

The velocity of the shock front decreases during the time because of adiabatic expansion and cooling of the bubble, it might happens the shock velocity equals the intergalactic gas sound speed, in this case, similarly to \citet{Bertone05}, we assume the dynamics of the wind joins the Hubble flow and no more mass is accreted. 

Since the Eqs.~\ref{dvdt}, \ref{dedt} and \ref{dTdt} requires an initial condition which cannot be $0$ (no bubble) to be consistently solved, we used the Sedov solution for a very short time with respect to $t_H(z=20)$ to obtain the \emph{primordial} bubble to follow the evolution. The evolutionary equations (Eqs.~\ref{dvdt}, \ref{dedt} and \ref{dTdt}) can be integrated numerically to follow the evolution of the shell along with the thermodynamic properties of the bubble for a given halo. 

A final point concerns the treatment of bubbles when two halos merge along the hierarchical merger tree. In that case, we impose mass conservation, i.e. the mass the new shell is the sum of the masses $M_s^{1}$ and $M_s^{2}$ of the two progenitor bubbles, that is $M_{s}^{f}=M_s^{1}+M_s^{2}$. Similarly, the final volume is taken to be the equal to sum of the volumes of the two single cavities $V_{b}^{f}=V_1^{f}+V_2^{f}$, and by the adopted spherical symmetry, we update the shock radius as $R_{s}^{f}= \left[ (R_s^{1})^3 + (R_s^2)^{3} \right]^{1/3}$. Finally, the shock velocity is given by the momentum conservation, which states that $v_s^{f} = \tilde{M}_s^{1} v_{s}^{1}+ \tilde{M}_s^{2} v_{s}^{2}$ where $\tilde{M}_s^{i} = M_s^{i}/(M_s^{1}+M_s^{2})$, and the new internal energy is the sum of the progenitor bubbles internal energies, that is $E_{b}^{f} = E_{b}^{1} + E_{b}^{2}$.  The gas temperature in the merged bubble is computed by assuming the final thermal pressure is the sum of the thermal pressures of the merging bubble: $T_{b}^{f} = \tilde{\rho}_b^{1} T_{b}^{1}+ \tilde{\rho}_b^{2} T_{b}^{2}$ where $\tilde{\rho}_b^{i} = \rho_b^{i}/(\rho_b^{1}+\rho_b^{2})$.

\subsection{Turbulence evolution}\label{Sub:turbulence}

In order to model the evolution of the turbulence developed in the expanding shells we adopt in the following an approach based on the spectral transfer equation derived by \citet{Norman96} based on the hydrodynamic Kovasznay approximation. In fact, our aim is to derive  averaged properties of the turbulent spectrum.

It is useful to introduce a spectral representation of the velocity field in the fluid. Then the energy density in eddies with wave-number between $k$ and $k+dk$ is given by $\rho \Spec(k,t) \de k$, where $\rho$ is the density of the fluid and $\Spec(k,t)$ is turbulence spectrum.
 
The time evolution of this quantity, under the assumption of local mode interaction in every region of the spectrum, is given by the following equation:
\begin{equation}\label{kova}
\frac{\partial}{\partial t} \Spec(k,t) = -\alpha \frac{\partial}{\partial k} \left[ \Spec(k,t)^{3/2} k^{5/2} \right] - 2 \nu k^2 \Spec(k,t) + \Source(k,t)
\end{equation}
where $\nu$ is the kinematic viscosity and $\Source(k,t)$ is the source function and $\alpha$ is a dimensionless constant that should be fixed experimentally. 

In order to describe the evolution of the turbulent spectrum, we have finally to provide the appropriate boundary and initial conditions:
\begin{gather}
\Spec(0,t) = \Spec(\infty,t) = 0 \, , \, 0 \le t < \infty \, ,\\
\Spec(k,0) = \Spec_0(k) \, , \, 0 \le k < \infty \, .
\end{gather}

A caveat must be made here. The assumption underlying the above description is one of incompressible, homogeneous, isotropic turbulence. This assumption clearly fails in the highly compressive IGM. However, the bulk of our knowledge on turbulence comes from terrestrial laboratory experiments, and most of them deal with liquids; in addition, simulations of compressible turbulence \citep{Lazarian04,Kowal07,Schmidt09} have shown that nonlinear interactions rapidly transfer most of the energy to non-compressible modes, in final one can use those results at least as a reasonable guide when discussing compressible turbulence.

To model $\Source(k,t)$ we assume that turbulent motions in the outflow are induced by interacting blast waves. That is, we maintain that observed turbulent motions in the shell are ultimately derived from the kinetic energy of SN induced shock waves which can act as source function for the turbulent cascade with an efficiency of order unity. If this is the case, there are some general constraints on the form of the source function. \citet{Bykov87} have studied in detail the shock-induced turbulence phenomenon, and conclude that the turbulent spectrum has, for the \citet{McKee77} model for the SN shock-wave expansions, a $k^{-2}$ dependence in the short-wavelength regime, while for long wavelengths it is proportional to $k^2$. Thus, the simplest source function retaining this behavior is
\begin{equation}
\Source (k,t) = S_0 (t) f(k/k_0)
\end{equation}
where $f(x)= x^2 / (1+x^4)$, $k_0$ is the wave-number corresponding to the characteristic length at which the turbulence is injected (and assumed to be $k_{0} \sim 1/R_{s}$) and $S_0(t)$ is a normalization factor obtained by equating the $k$-integral of the source function to the kinetic energy rate ($\dot{E}_{\rm kin}$). Finally, as in \citet{Branderburg07} we use a constant kinematic viscosity of $\nu = 5 \times 10^{24}$~cm$^2$~s$^{-1}$; this is equivalent to scale the dynamic viscosity coefficient with density.

The spectrum evolution is followed for each halo of the MT; again, we need to specify how we treat merger events. To conserve turbulent energy associated to any MT node we assume
\begin{equation}
\rho \Spec(k) = \rho_1 \Spec_1(k) + \rho_1 \Spec_2(k),
\end{equation}
which entails 
\begin{equation}
\Spec(k) = \frac{M_1 R_s^3}{M R_{s1}^3} \Spec_1(k) + \frac{M_2 R_s^3}{M R_{s1}^3} \Spec_2(k).
\end{equation}
The turbulent pressure is finally given by 
\begin{equation}\label{pturb}
p_{t}(t) = \rho \int_0^\infty \de k \Spec(k,t). 
\end{equation}


\section[]{RESULTS}

In this Section we present the results of our model. After a description of the model calibration, useful to fix the model free parameters, we give our prediction for the  bubble evolution along a MT extending down to $z=0$ along with the IGM turbulence properties. 

\subsection{Model calibration}\label{Sub:MW}
\begin{figure}
\centerline{\epsfig{figure=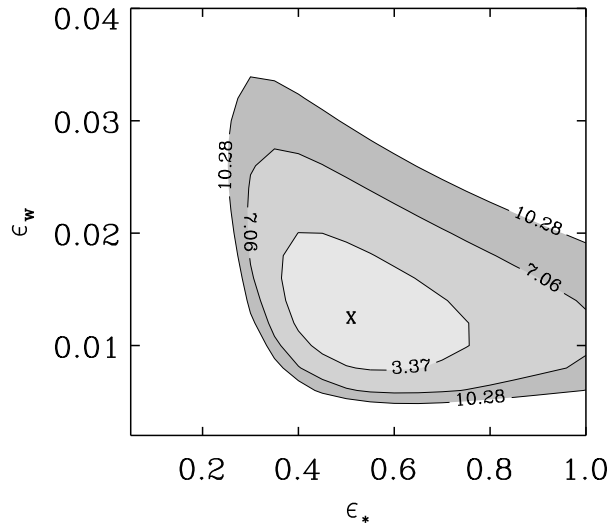,width=8cm,angle=0}}
\caption{The 68\%, 95\%, 99\% confidence level regions of our galaxy model are represented in the ($\epsilon_{\rm w}-\epsilon_*$) plane.}
\label{MW_CL}
\end{figure} 
\begin{figure}
\centerline{\psfig{figure=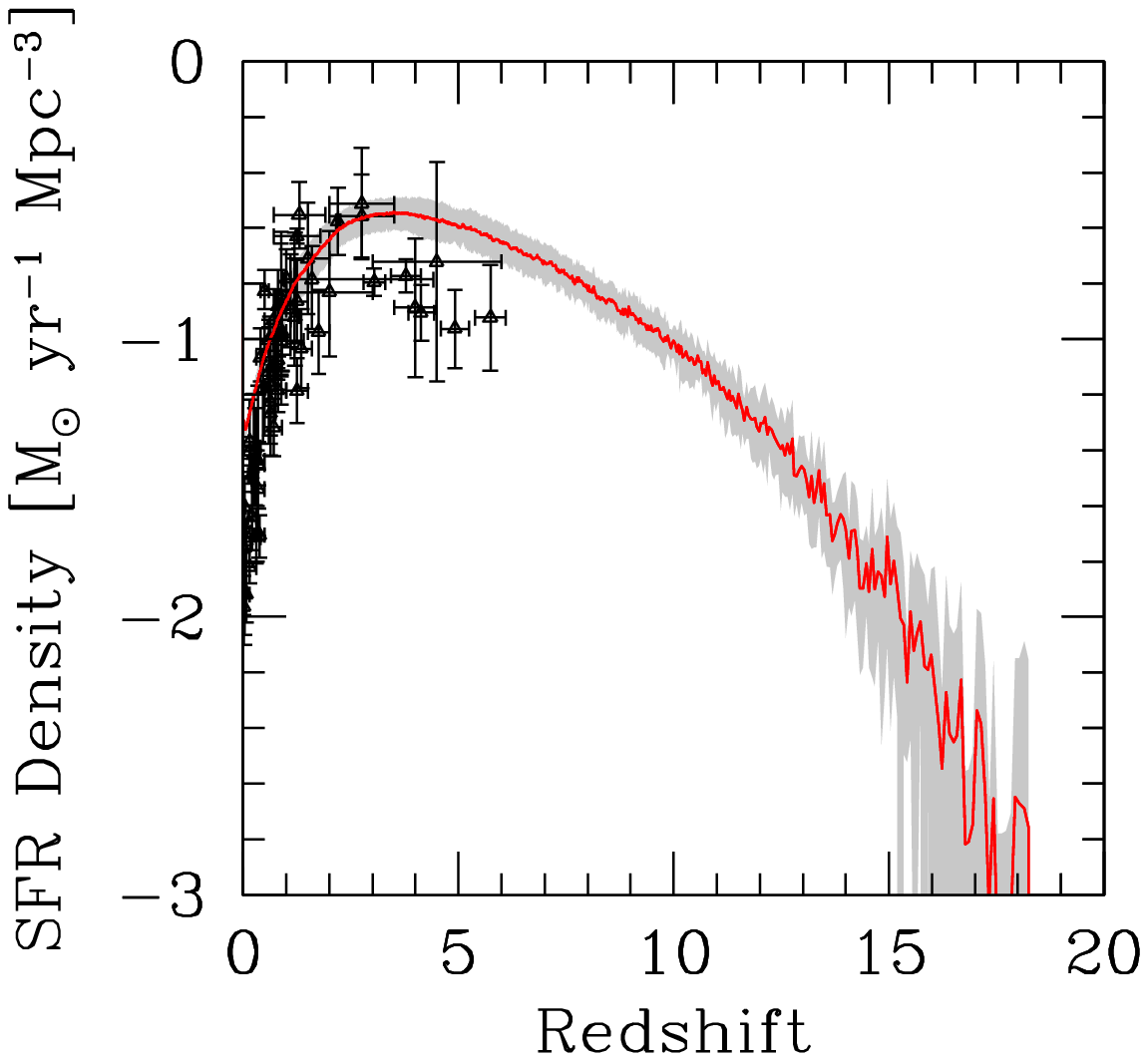,width=8cm,angle=0}}
\centerline{\psfig{figure=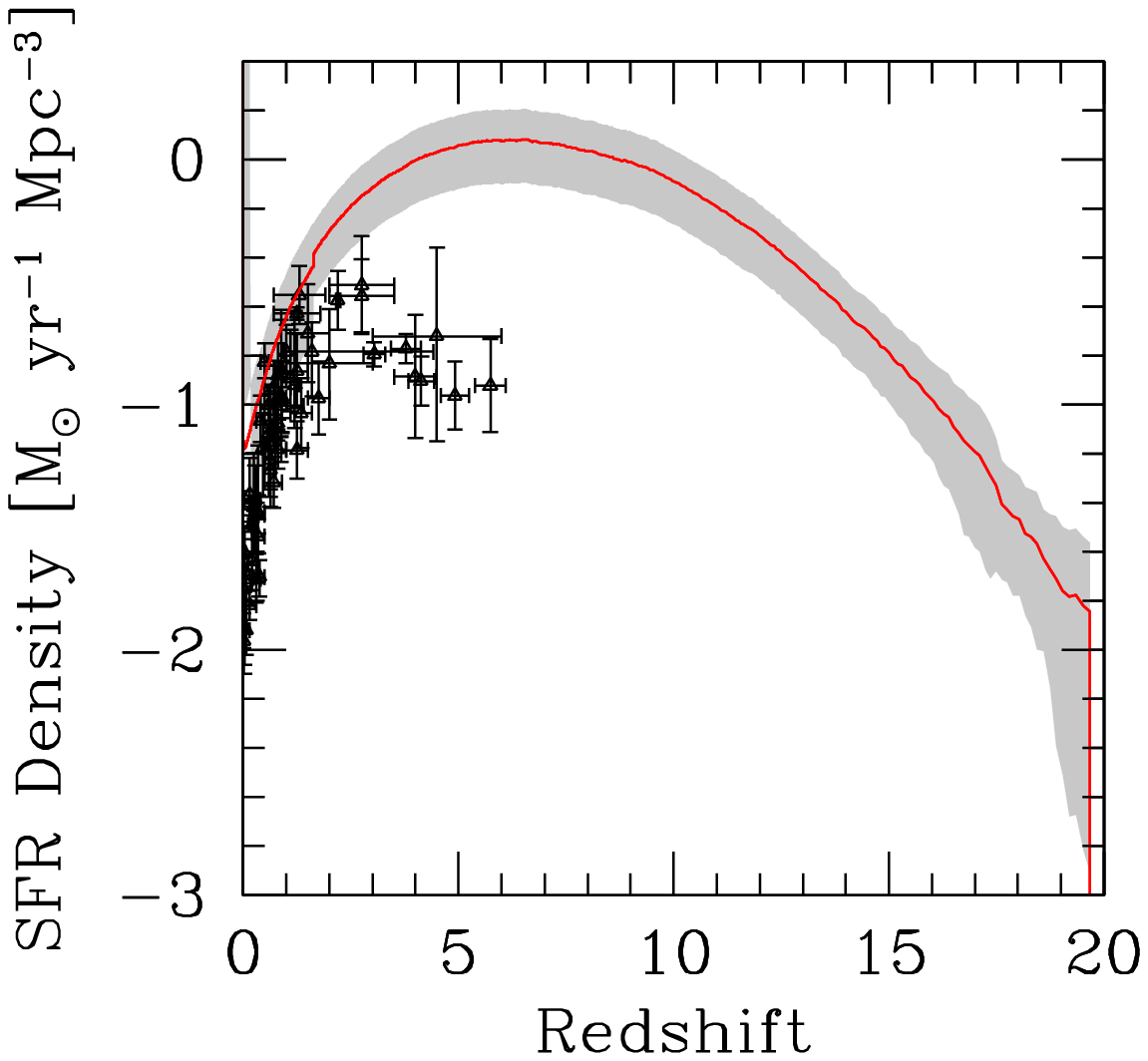,width=8cm,angle=0}}
\caption{Redshift evolution of the SFR density along the MW merger tree: (top) for our best-fit model ($\epsilon_* = 0.5, \epsilon_{\rm w}=0.012$), (bottom) for the no-feedback model ($\epsilon_* = 0.5, \epsilon_{\rm w}=0$). The curves are obtained after averaging over 100 realizations of the MT; shaded areas denote $\pm 1\sigma$ dispersion regions around the mean. Points represent the low-redshift measurements of the cosmic SFR by \citet{Hopkins06}.}
\label{MW_SFR}
\end{figure} 
Our galaxy evolution model includes several relatively poorly known (albeit important) physical processes that need to be empirically calibrated. To this aim we have used the observed properties of the MW as a benchmark to fix the best values for the two model free parameters: $\epsilon_*$, the efficiency of star formation (Eq.~\ref{sfr}) and $\epsilon_{\rm w}$, the SN feedback efficiency (Eq.~\ref{eje}). This has been accomplished by producing the MT of a dark matter halo with mass consistent with the MW one \citep[$10^{12}$~M$_\odot$,][]{Xue08} at redshift $z=0$. We have then compared the resulting properties of the synthetic galaxy with the following ones deduced from MW observations:

\begin{itemize}

\item Stellar mass. Contributions to the stellar mass come from the disk $M_*^{disk} \sim (4-6) \times 10^{10} M_\odot$, the bulge $M_*^{bulge} \sim (0.4 - 1) \times 10^{10} M_\odot$, and the halo $M_*^{halo} \sim (0.2 - 1) \times 10^{10} M_\odot$ components \citep{Dehnen98,Brown05}, yielding a total stellar mass of $M_* \sim 6 \times 10^{10} M_\odot$.

\item Gas-to-stellar mass ratio, $M_{gas}/M_* = 0.13$. The mass of gas has been derived using the observed mass of HI and HII regions of the Galaxy, $M_{gas} = M_{HI} + M_{HII} \sim (6+2) \times 10^9 M_\odot$ \citep{Stahler05}.

\item Current star formation rate. \citet{Murray10} use the total free-free emission in the WMAP foreground map as a probe of the massive star population and derive a global SFR of $1.3$~M$_\odot$~yr$^{-1}$.

\end{itemize}

For all these observables we assume a relative uncertainty of $20\%$.
Fig.~\ref{MW_CL} shows the $\chi^2$ confidence levels of our model with respect to these observables as function of $\epsilon_*$ and $\epsilon_{\rm w}$. The distribution presents a clear minimum at ($\epsilon_*,\epsilon_{\rm w}$) = (0.5, 0.012); however, a degeneracy exists with a relatively wide range of choices that can reproduce with sufficient accuracy the global MW properties. Thus, the best-fit model implies a star formation timescale which is only a factor $\approx 2$ longer than the free-fall time and a relatively inefficient feedback in order not to expel from very first proto-galaxies too much gas which will then be crucial to fuel the subsequent SF in larger objects.

From the same run we can obtain the redshift evolution of the SFR density along the MW merger tree for our best-fit model, after averaging over 100 different realization of the MT\footnote{The effective comoving volume of the MW is taken equal to $30$~Mpc$^3$, and corresponds the size of a $10^{12}$~M$_\odot$ linear fluctuation.}. For comparison sake only, we overplot the cosmic SFR on top of the MW SFR history. Perhaps not coincidentally, the two evolutions match each other quite well indicating that the Galaxy is a not too biased tracer of average cosmic conditions. Finally, in the same figure we show the SFR density evolution for a model with $\epsilon_{\rm w} = 0$; you can see that in this case the model is far from the data, which means that, even if the efficiency happens to be small, the feedback by galactic winds is crucial to obtain a consistent model of galaxy evolution.

In the remainder of the paper we assume that these values for ($\epsilon_*,\epsilon_{\rm w}$)  hold for any galaxy present in the MT. This might a poor approximation, but given the persisting ignorance on star formation and feedback processes  a better performance of more complex choices is not guaranteed to produce a more solid result. In any case the model success in reproducing key Galactic properties provides at least a first and necessary step towards more refined treatments. 
\subsection{Galactic outflows}\label{Sub:out}
\begin{figure*}
\center{\psfig{figure=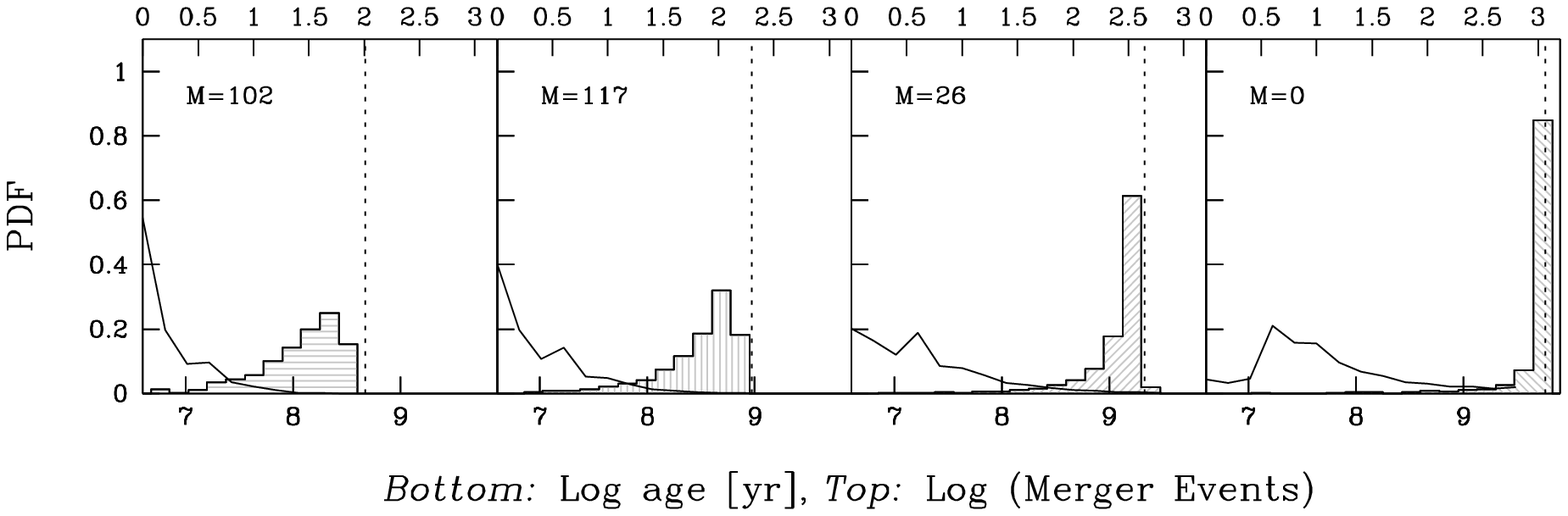,width=15cm,angle=0}}
\center{\psfig{figure=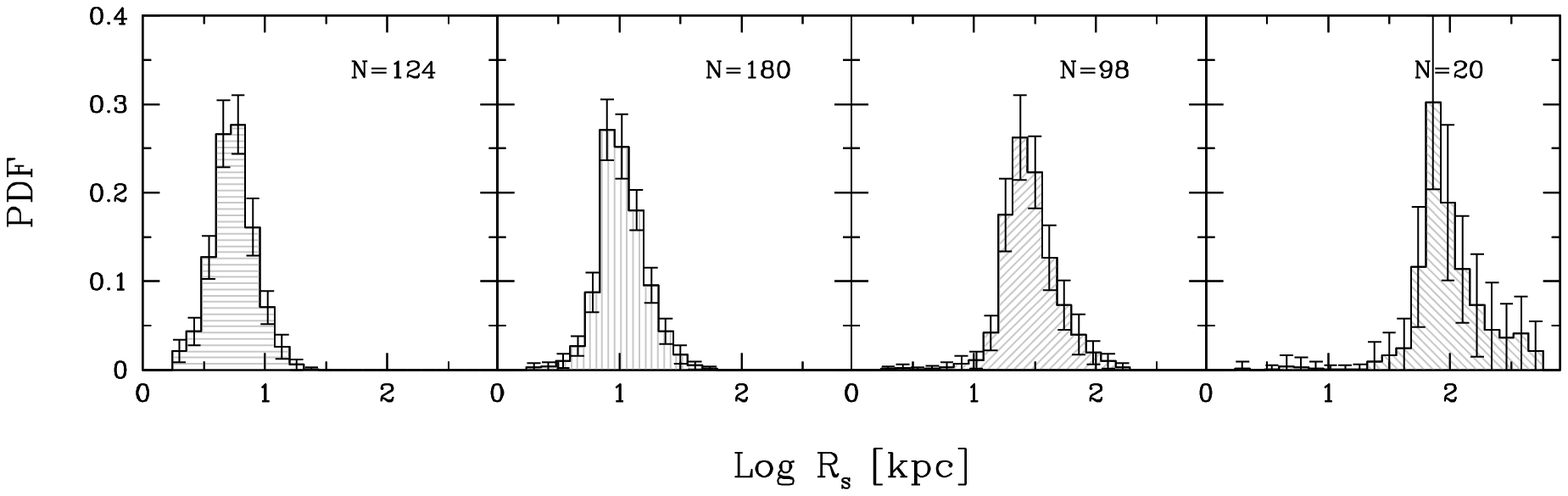,width=15cm,angle=0}}
\center{\psfig{figure=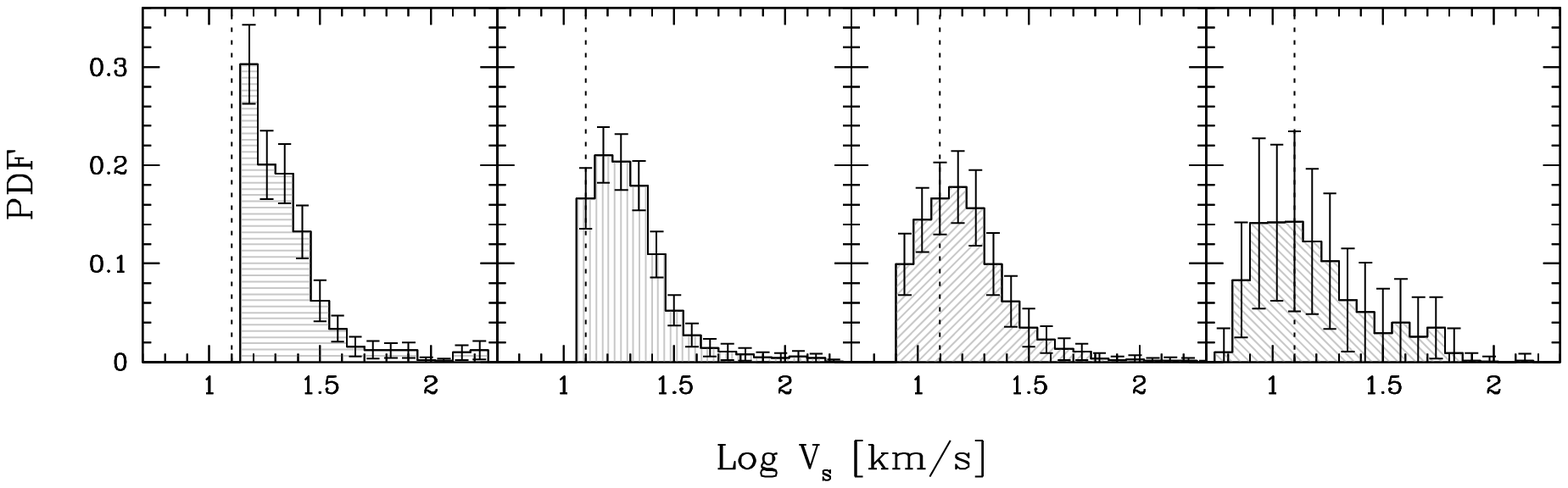,width=15cm,angle=0}}
\center{\psfig{figure=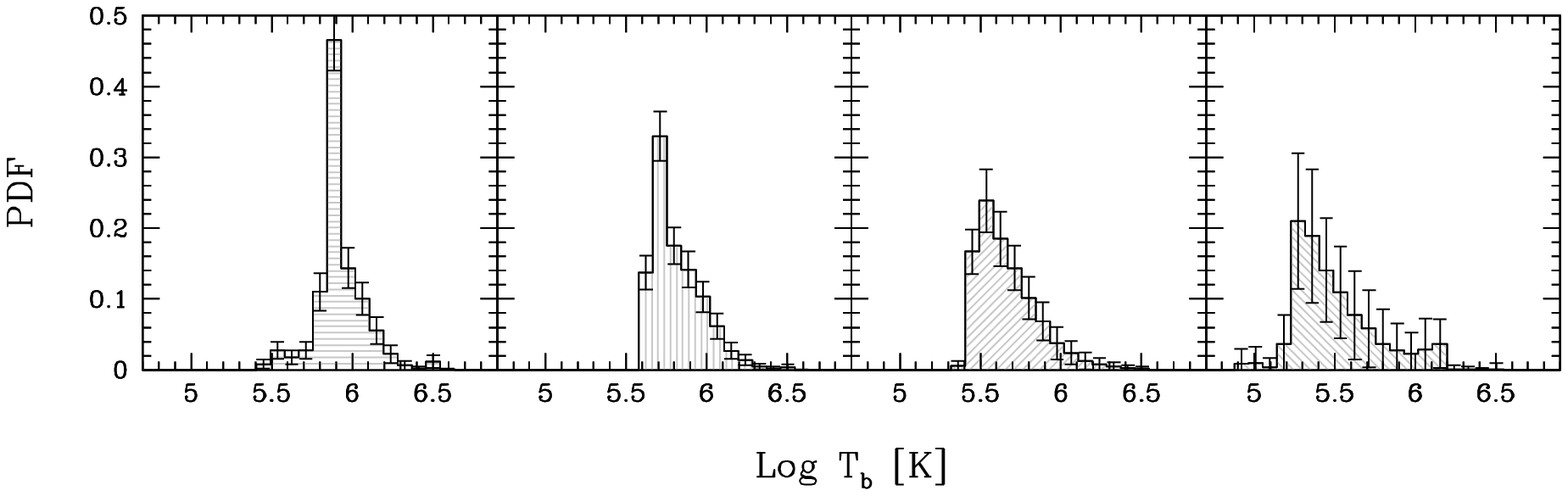,width=15cm,angle=0}}
\caption{Probability Distribution Function (PDF) for different bubble properties (age, bubble radius $R_{S}$, shell velocity $V_{S}$ and internal temperature $T_{b}$) at different redshifts (left to right) $z=10,6,3,1$; error bars show the $1\sigma$ dispersion. The solid curves in the top panels show the distribution of the number of mergings for the halos in the MT. The label $M$ indicates the number of halos that evolved in isolation up to that redshift; $N$ indicates the total number of halos at that $z$. The dashed lines in the top panels marks the age of the universe at that redshift. The dashed lines in the third row panels marks the IGM sound speed at that redshift.}
\label{bubbles}
\end{figure*} 
We illustrate the properties of galactic outflows by focusing on the case of the progenitors of a $10^{13}$~M$_\odot$ halo at redshift $z=0$ (i.e. a small cluster/group). Since the total number of galaxies at any intermediate redshift is very large we will deal with \emph{mean} quantities to describe the global properties of the winds. This precludes the inspection of one-to-one relationships between galaxy and its wind properties, but will make possible to appreciate the correlations among several global quantities, as we will see in the following. 

Fig.~\ref{bubbles} gives an concise overview of the bubbles we have identified around galaxies at selected redshifts ($z=10,6,3,1$) and it can be used to elucidate many physical aspects. Bubbles tend to become older towards lower redshifts with a decreasing age spread; their ages tend to accumulate close to the Hubble time. This behavior reflects the fact that after an early evolutionary phase in which bubbles grow by number around relatively low-mass objects, the average growth of the halo population mass associated with deeper potential wells, prevents the formation of new bubbles after $z=3$. Hence most of the bubbles seen around galaxies at intermediate redshifts ($z=3-5$), where they are more easily observed, contain material expelled by galaxies at a much earlier time.  At the same time old bubbles merge together to form a fewer large ones. The role of merging events is made clear by the solid curves in the top panels which show the distribution of merging events for halos that have witnessed at least 1 merging event. A fraction $M/N$, indicated by the labels inside the panels in Fig. \ref{bubbles}, have evolved in isolation until the considered $z$, i.e. they never suffered a merger. Such ratio decreases steadily with time, going from $81$\% at $z=10$ to $0$ at $z=1$. The merging activity increases the average ages of bubbles: as discussed above, at high $z$ a large number of young, isolated bubbles exist that are later turn into aged, large ones.

The typical bubble size grows from 5~kpc (physical) at $z=10$, to about 100~kpc at $z=1$. Merging is definitely an important agent behind this evolution; however, it cannot be the only growth factor for bubbles. This can be realized by combining the information from the age and size distributions. At the lowest redshifts shown in Fig. \ref{bubbles}, the bubble age distribution essentially collapses onto the cosmic age value; nevertheless, a considerable size spread, ranging from $1$ to a few hundred kpc, exists. Such spread is caused by the corresponding dispersion of galaxy star formation rates. Larger galaxies tend to have faster winds, which show up as a tail in the PDF of the shock velocity distributions ($V_s > 30$ km s$^{-1}$). Hence a minority of the bubbles found at $z<3$ are constituted by old bubbles which are still powered by their current galaxy, which is undergoing vigorous star-formation. Instead of creating a new bubble, these massive cluster progenitors blow their winds into pre-existing, old bubble structures. 
 
From the velocity PDF evolution, it is interesting to note that at early times bubbles (somewhat contrary to naive expectations) expand on average at larger velocities and they tend to decelerate at lower redshifts. This is an obvious implication of the aging trend described above. As larger galaxies dilute their energy on the larger volume of pre-existing cavities, they can sustain the growth of bubbles but cannot re-accelerate them to the high velocities at which they were traveling in the past.

The evolution of the bubble temperature $T_b$ (Fig.~\ref{bubbles}) is governed by competing effects. Adiabatic expansion and radiative cooling (depending on bubble gas temperature and, to a much limited extent, gas metallicity) lead to a decrease of $T_b$, while thermalization of the wind energy provide a heat input. Almost all the bubbles have temperatures $> 10^{5.5}$ K at $z=10$, whereas by $z=3$ this fraction decreases to $< 20$\%. This is mainly due to the adiabatic expansion of the bubbles.
\begin{figure}
\center{\psfig{figure=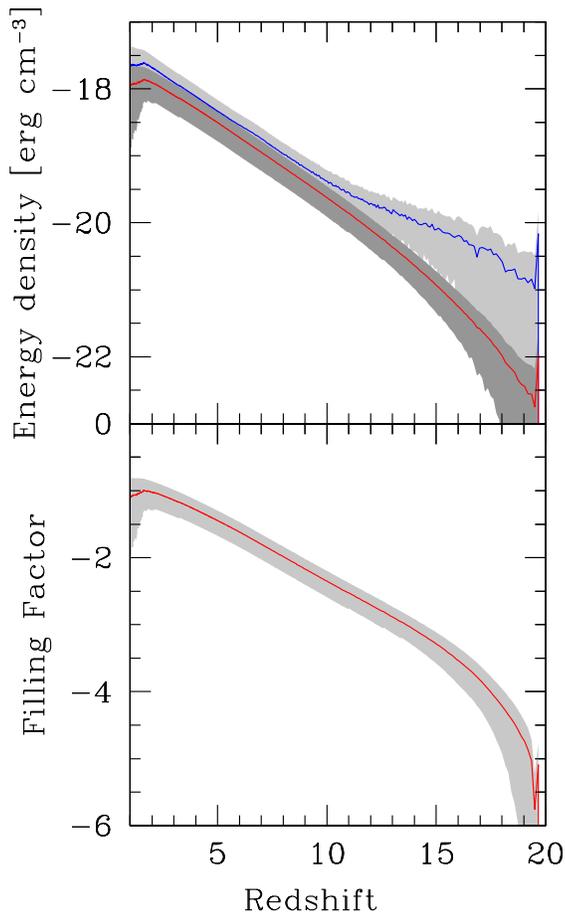,width=8cm,angle=0}}
\caption{
\emph{Top:} Mean kinetic (blue) and thermal (red) energy of regions affected by outflows at different redshifts. 
\emph{Bottom:} Volume filling factor of outflows. Shaded areas show $1\sigma$ dispersion around the mean (solid curve) of 100 merger tree realizations.}
\label{benergy}
\end{figure}
The expanding outflow is powered by the continuous injection of SN energy both in kinetic (in practice stored in the shell, where most of the mass is located) and thermal (stored in the bubble interior) forms. More quantitatively, from Fig.~\ref{benergy} one can appreciate that the kinetic energy is always dominant (by a factor 2-3) with respect to the internal one. This kinetic energy will be rapidly converted in disordered motions by instabilities, finally resulting in a fully developed turbulent spectrum. The bubble volume filling factor also increases with time, as illustrated in the bottom panel of the same Figure, reaching about 10\% at $z < 2$. One has to keep in mind that our study is based on the MT of a cluster/small group and therefore it might reflect a somewhat biased cosmic regions, although we do not expect major differences as we extrapolate these results to general IGM. A similar treatment of the outflow evolution has been successfully used by \citet{Bertone05} and \citet{Samui08} to investigate the role of galaxies in enriching the IGM. In particular \citet{Samui08} use the modified PS formalism to follow the evolution of halos and porosity-weighted averages to investigate the global influence of winds. For a reasonable choice of the model free-parameters, they find that outflows can generally escape from the low mass halos ($M_h < 10^9 \, M_\odot$) which then dominate the observed enrichment; nevertheless, MW-type galaxies can create Mpc-size metal enriched bubbles in their surroundings. \citet{Bertone05} apply the semi-analytical prescription for galactic winds to high-resolution N-body simulations of field galaxies. Their main conclusion is that galactic outflows do not perturb the structure of the Ly$\alpha$ forest, galaxies with $10^9 < M_* < 10^{10}$~M$_\odot$ are the main responsible for IGM chemical enrichment at $z=3$. 

The bubble volume filling factor is a useful tool when applying our models to quantify turbulent effects in the quasar absorption line spectra, as we will try to do later on. As a note of caution, the filling factor shown in Fig.~\ref{benergy} might be overestimated to some degree as we do not attempt to account for possible, although likely rare, overlaps between bubbles.  


\subsection[]{Turbulence evolution}\label{Sub:oturb}

The best way to observationally probe the turbulent content of gas affected by outflows is to measure the Doppler parameter, $b$, of absorption lines arising from kinetically perturbed gas. For this reason in the following we will use $b$ to quantify turbulence evolution. 

In general, one can write $b$ as the quadratic sum of the thermal and turbulent components,   
\begin{equation}
b = (b_{\rm th}^2 + b_{\rm t}^2)^{1/2},
\end{equation}
where 
\begin{equation}
b_{\rm th} = \left( \frac{2k_B T}{m_a} \right)^{1/2}
\end{equation}
is the thermal contribution and 
\begin{equation}\label{pressure}
b_{\rm t} = \sqrt{ \frac{p_{\rm t}}{\rho} }
\end{equation}
is the contribution due to the turbulent pressure defined in Eq.~\ref{pturb}. Note that the turbulent contribution (contrary to the thermal one) does not depend on the mass $m_a$ of the atomic mass of the absorbing element. This allows to generalize our results to all species associated with the cold HI shell gas (e.g. CIV). Moreover, since the observed Doppler parameter is the sum of the two contributions and the thermal broadening is smaller for heavier elements, the relative contribution of turbulent broadening should more important (and hence more easily detectable) for heavier elements.
\begin{figure*}
  \center{\psfig{figure=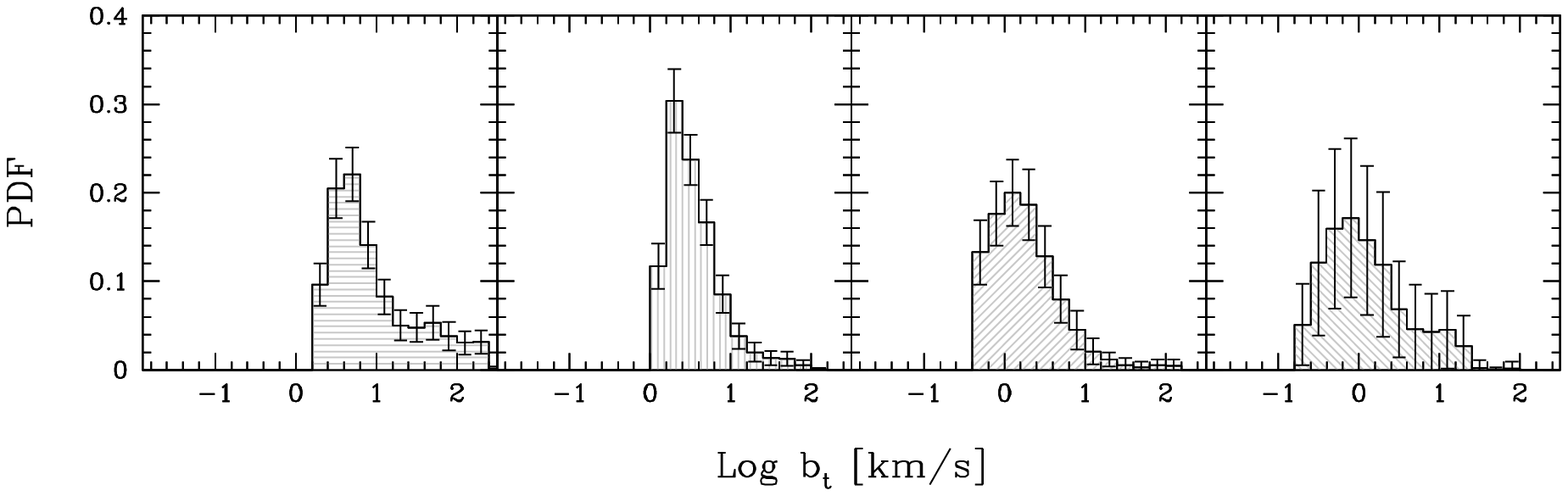,width=15cm,angle=0}}
  \center{\psfig{figure=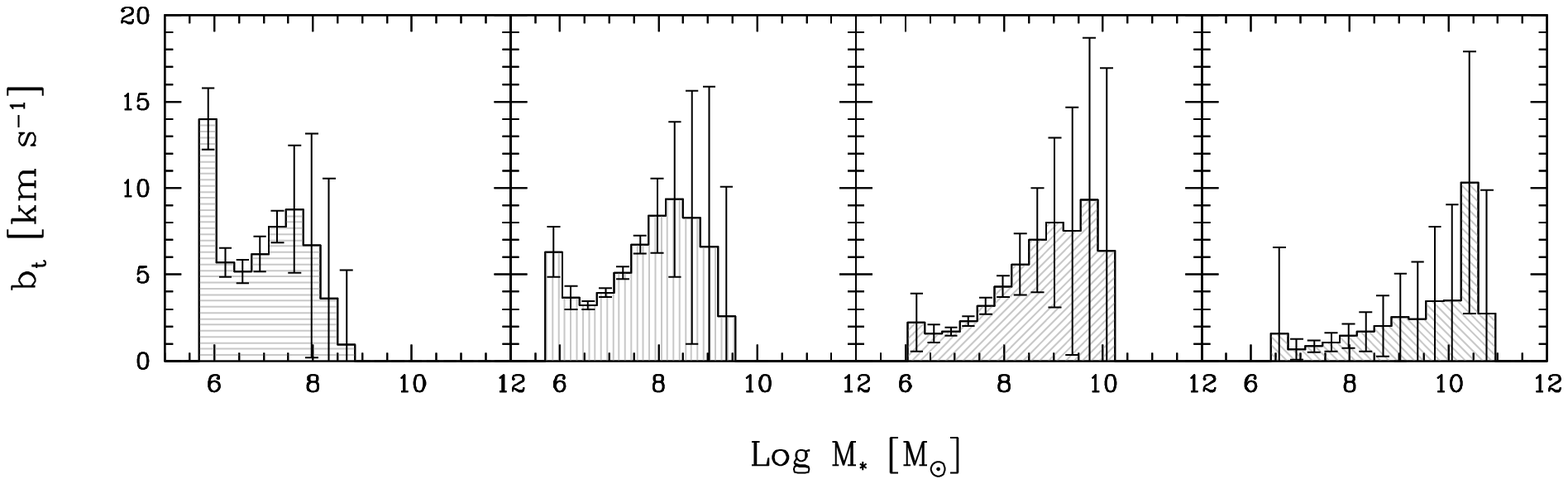,width=15cm,angle=0}}
  \center{\psfig{figure=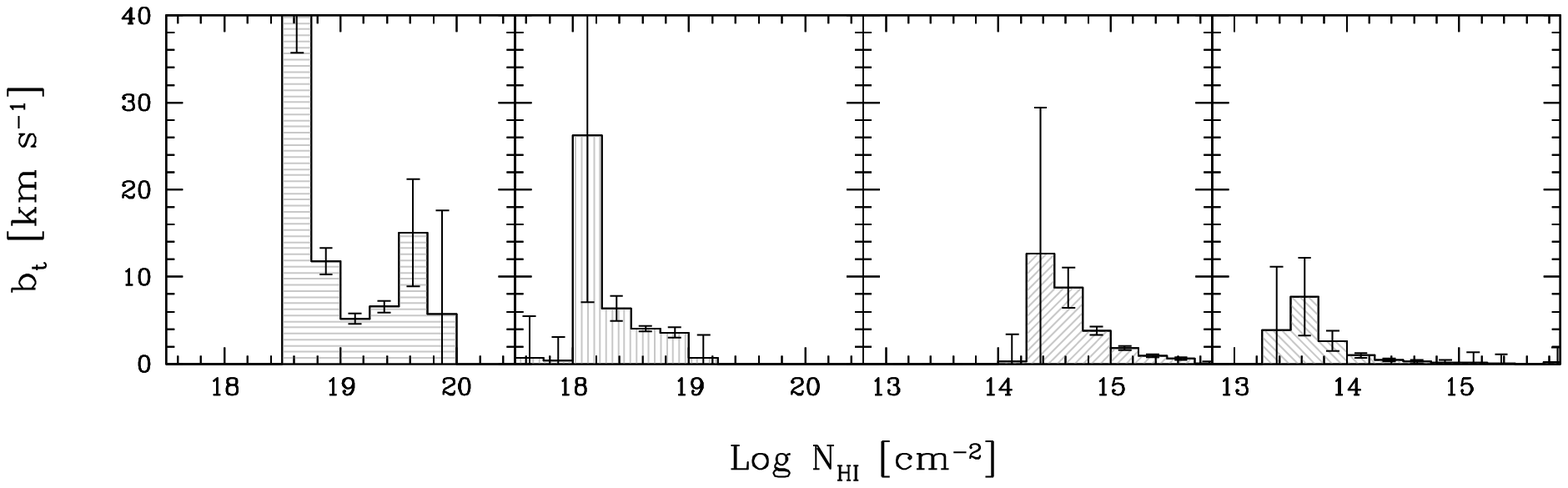,width=15cm,angle=0}}
  \center{\psfig{figure=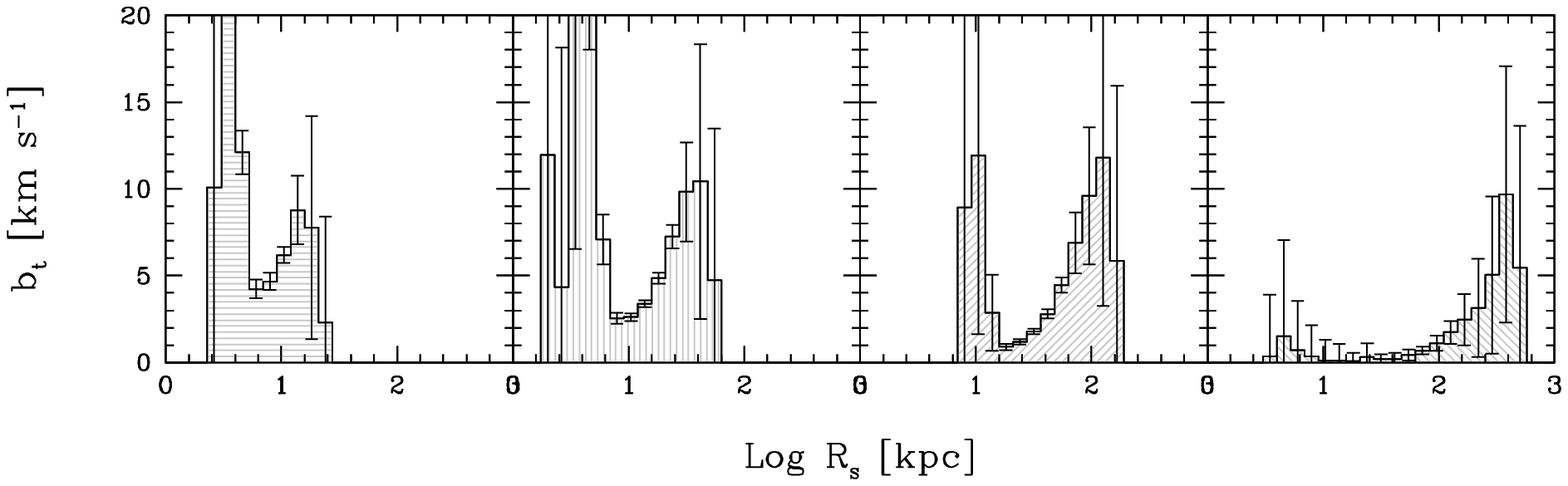,width=15cm,angle=0}}
  \caption{Turbulence properties in terms of the Doppler parameter $b_t$ at four selected redshifts:  $z=10,6,3,1$ (left to right columns). The first row contains the PDF of $b_t$, the others (top to bottom) show the relations between $b_t$ and stellar mass, $M_*$,  hydrogen column density, $N_{HI}$, shock radius, $R_s$, respectively. Error bars denote $1\sigma$ dispersion in each bin.}
\label{turbs}
\end{figure*} 

In Fig.~\ref{turbs} we show the distribution of Doppler parameters in simulated galaxies at different redshifts. The $b_t$ distribution function at all redshifts shows a similar shape, with a low-energy cut-off, followed by a maximum and a steed declining tail towards larger values. However, the PDF shift towards lower $b_t$ values with time: while at $z\ge 6$, the region affected by the winds can become very turbulent, reaching in a few spots $b_t > 50$~km~s$^{-1}$, as a result of the violently expanding of fresh new bubbles carved by early galactic winds, at later times winds blowing in pre-existing bubbles limit become more gentle, producing in decrease of turbulent broadening effects. By $z\approx 1-2$ the distribution has stabilized on median values of the order of 1-2 km~s$^{-1}$, and even the most turbulent spots do not exceed 25 km~s$^{-1}$. The trend above is further enhanced by the dissipation of turbulent energy with time, which does not allow to store kinetic energy into eddies indefinitely. 

To see this quantitatively, it is useful to estimate the turbulent dissipation timescale. Under the hypothesis of a fully developed, steady-state spectrum the dissipation timescale is well approximated by the eddy turn-over timescale, namely $t_e \sim R_s/V_s$. If we further assume that to the early stages of bubble evolution the Sedov solution applies, it follows that $t_e \sim 2.5 t_b$, where $t_b$ is the bubble age, in turn shorter than the Hubble time. This implies that, due to their high velocity and relatively small radii, dissipation acts very efficiently at earlier stages, thus decreasing the turbulent energy budget of these objects.
\begin{figure}
  \centerline{\psfig{figure=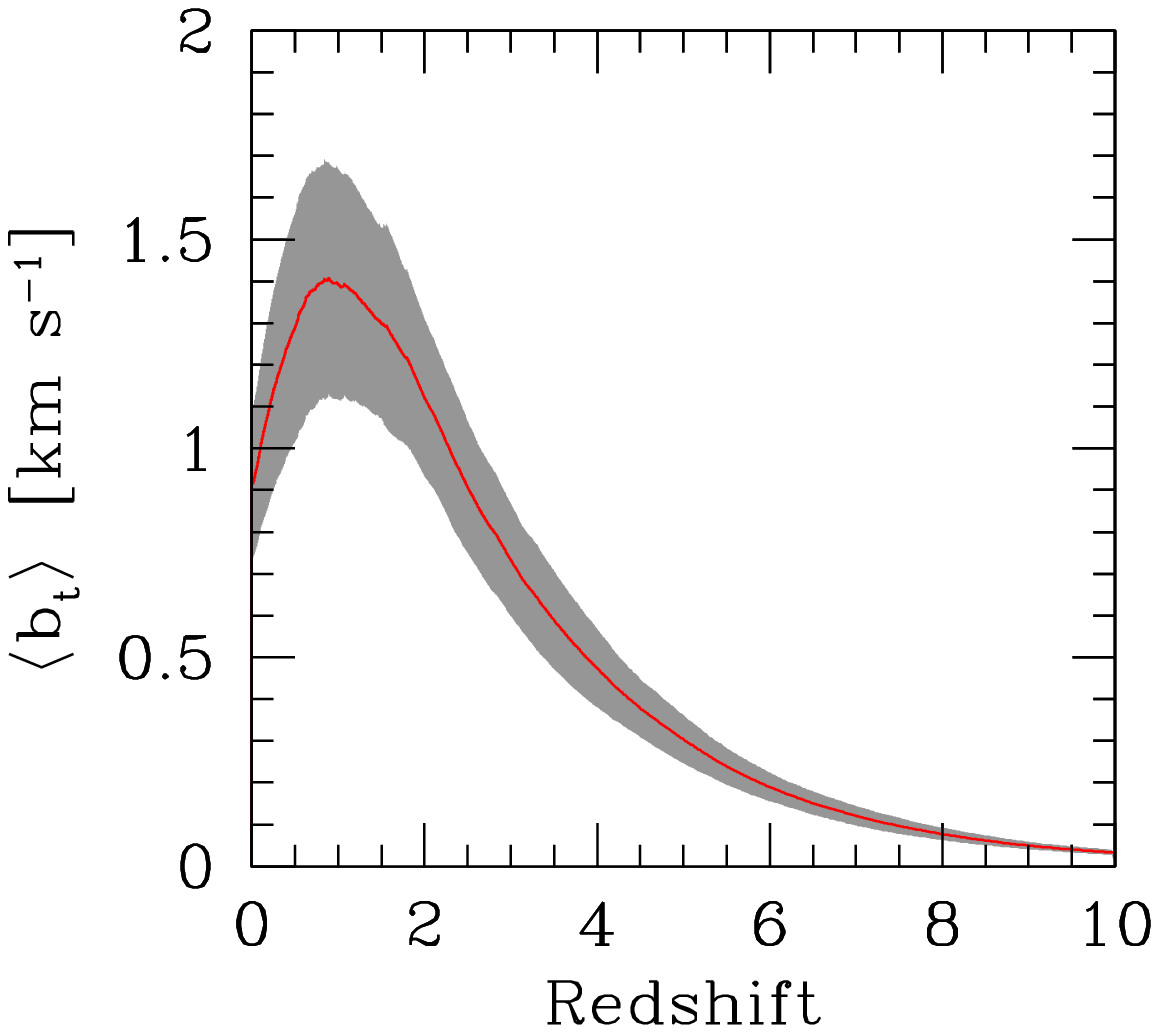,width=8cm,angle=0}}
  \caption{Redshift evolution of the volume-weighted Doppler parameter. The solid curve shows the mean; the shaded area corresponds to $1\sigma$ dispersion.}
  \label{VW_BP}
\end{figure} 

We pause briefly to emphasize an important point. One has to keep in mind though (see Fig. \ref{benergy}) that the filling factor of the turbulent regions is quite small at high redshift, and therefore it must be made clear that the above is not the $b_t$ parameter applicable to the global IGM, but only to tiny and specific (but well defined) portions of it. By plotting the volume-weighted evolution of $\langle b_t \rangle$, shown in Fig. \ref{VW_BP}, we can find support to the previous statement: at higher redshift the turbulent pressure contribution is less important because of the small size of outflow-affected regions, whereas in local objects the bubbles grow enough to occupy a significant fraction of the volume; however, this occurs too late, when dissipation has already started to quench eddies on the smallest scales. The trade-off between injection and dissipation, conspire to produce a maximum in $\langle b_t \rangle$ at $z \approx 1$.  
\begin{figure}
  \center{\psfig{figure=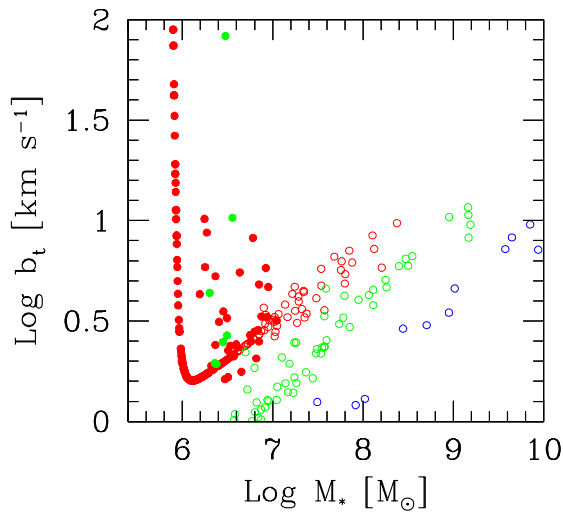,width=8cm,angle=0}}
  \caption{The Doppler parameter $b_t$ as a function of the galaxy stellar mass for a single realization of the MT, shown at $z=6$ (red circles), $z=3$ (green circles), $z=1$ (blue circles). The filled points indicate bubble whose age is less than half of the Hubble time.}
  \label{bp_vs_mstar}
\end{figure}

Let us turn back to the analysis of the other panels in Fig.~\ref{turbs}. It is interesting to determine the level of turbulence in bubbles produced by galaxies of different stellar masses. In general, $b_t$ is positively correlated with the stellar mass of a galaxy. This is a consequence of the fact that the SFR (which is ultimately proportional to the energy input rate due to supernovae) is more sustained in larger galaxies. However, we see that the onset of a turbulent regime is almost unavoidable in regions that are stirred by galactic winds.   
Again, the peak at very low $M_*$ is interpreted as the imprint of young bubbles in which turbulence has not yet suffered from dissipation: we double-checked this hypothesis by studying the correlation in different bubble age bins. This analysis is reported in Fig. \ref{bp_vs_mstar} (for a single MT realization) whose examination clarifies that the youngest bubbles have the largest $b_t$ values. Thus, the bursting star formation activity going on in small galaxies momentarily reverses the trend of increasing $b_t$ as a function of stellar mass, clearly indicating that the level of turbulence depends both on age {\it and} mass of the object. The spread of the distribution is very limited, suggesting that different mass accretion histories do not influence substantially final Doppler parameter distribution.

A relatively easily accessible observational quantity when studying the IGM is the neutral hydrogen column density, $N_{HI}$, of the absorbing shell/system. A reasonable estimate for such quantity can be obtained from the following formula:
\begin{equation}
N_{H} = \int_{R_{s}-\Delta R_{s}}^{R_{s}} n_{H}(r) dr \approx n_{H} 
\Delta R_{s} = \frac{M_{s}}{4\pi \mu R_{s}^{2}} 
\end{equation}
In the latter $n_{H}$ is the mean density in the thin shell, given by the ratio between its mass ($M_{s}$) and its volume $V_{s}$:
\begin{equation}
V_{s} = 4 \pi R_{s}^{2} \Delta R_{s}
\end{equation}
The neutral hydrogen ionization correction can be obtained by solving the appropriate photoionization equilibrium balance equations given, e.g. by \citet{Theuns98} and assuming the UV background history of the Universe computed by \citet{Haardt96}. The third row of panels in Fig. ~\ref{turbs} shows the distribution of $b_t$ with $N_{HI}$. One can note that at redshift $z\sim 3$ the turbulent broadening starts making an impact in HI lines with $N(HI) \geq 10^{14}$~cm$^{-2}$.

Finally, the lowermost panel in Fig. ~\ref{turbs} shows the correlation between $b_t$ and the distance reached by the outflow. This might have important observational implications. \citet{Adelberger03} by studying the correlation between galaxies and CIV systems with large velocity spreads, showed that most of these systems are associated with star-forming galaxies. It is conceivable to interpret such result as a direct evidence of large-scale outflows around $2\lsim z\lsim 3$ galaxies. A similar investigation, as discussed in the next Section, could directly probe the level of turbulence in these objects and test our predictions.

\section{Test the model against data.}

\begin{figure*}
  \center{\psfig{figure=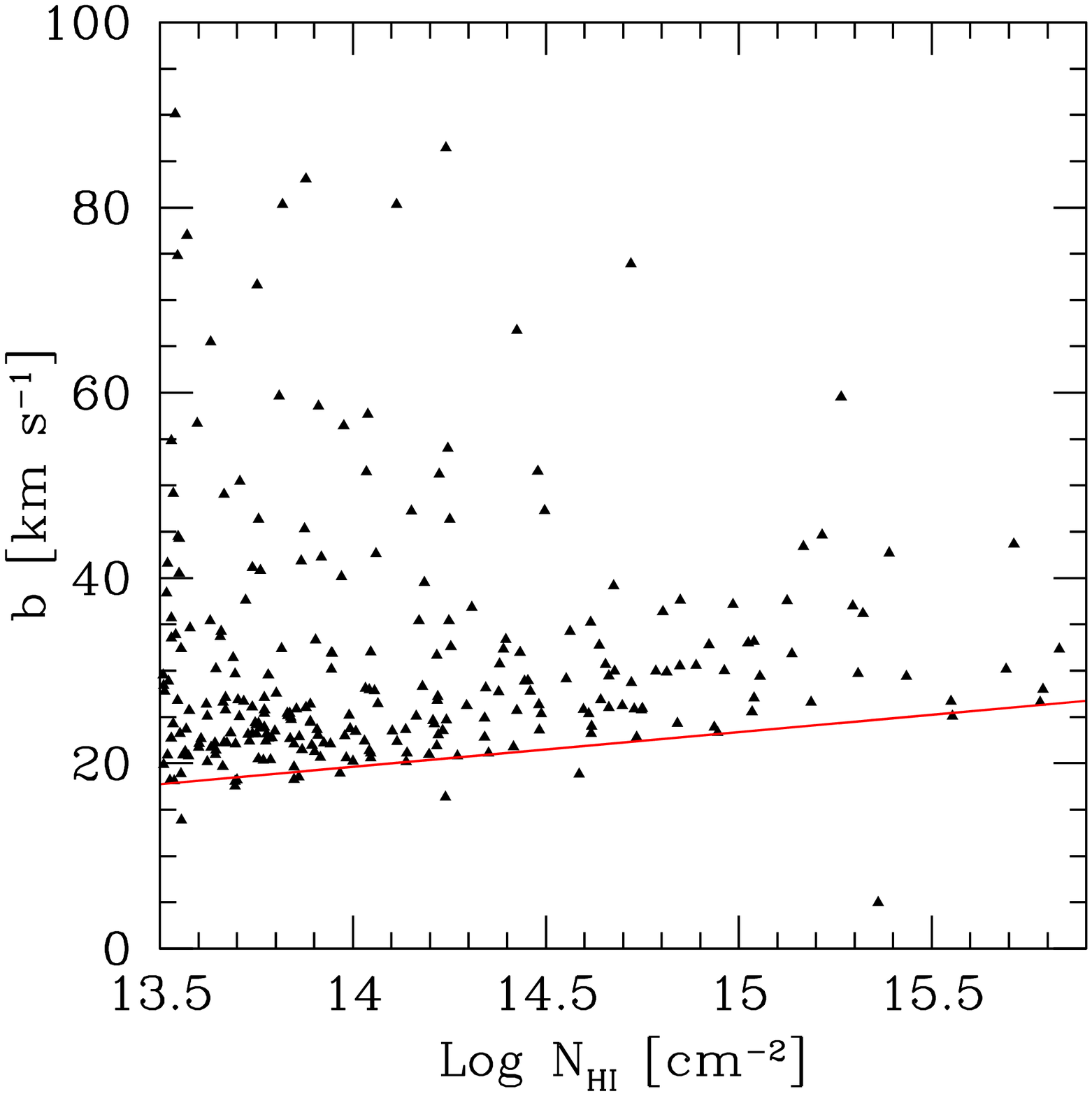,width=8cm,angle=0}
          \psfig{figure=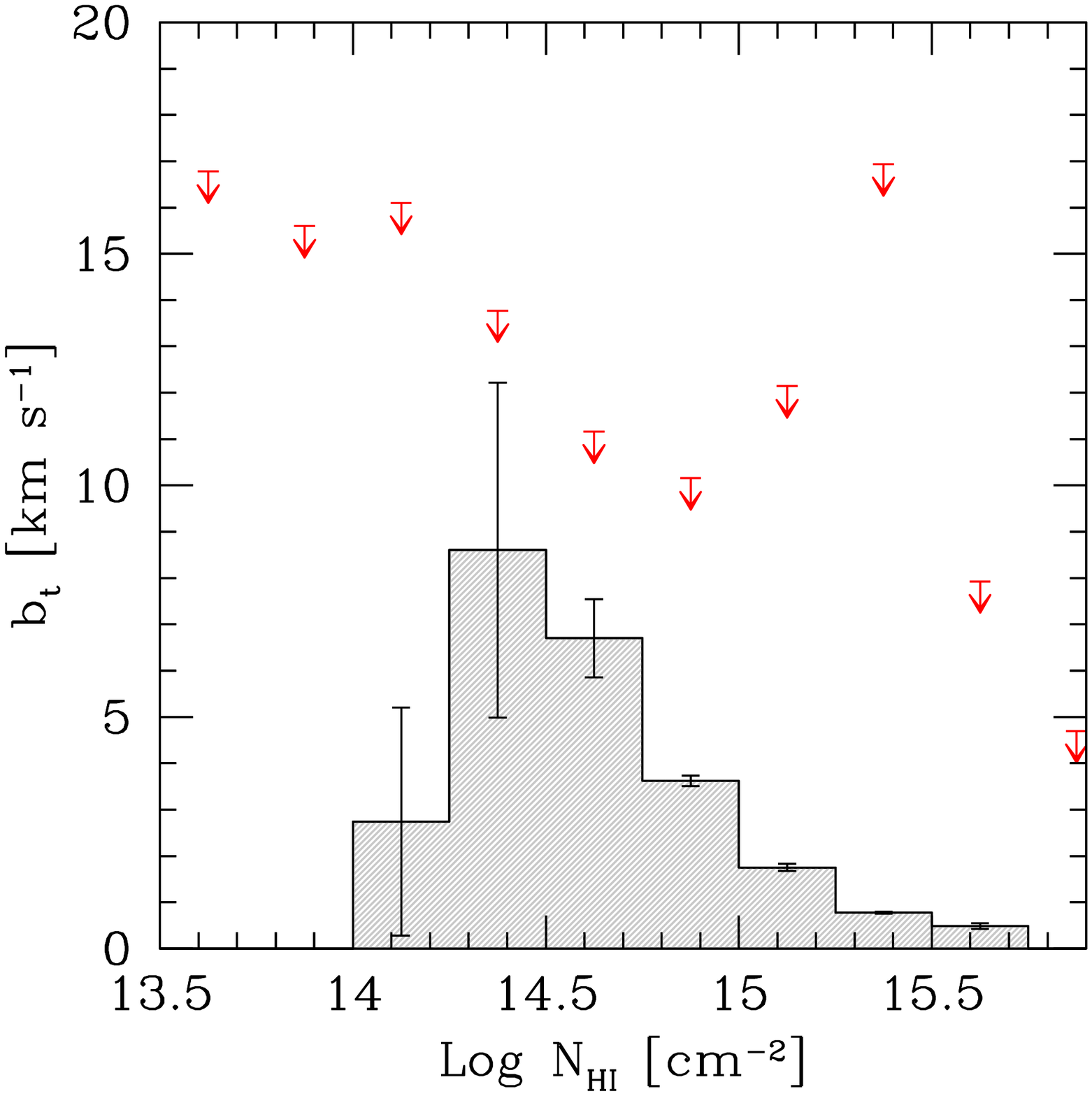,width=8cm,angle=0}}
  \caption{{\it Left panel:} Doppler parameter -- HI column density relation observed in the IGM at $z=3$ (data points from VLT/UVES LP). The red curve represents a fit to the lower cutoff used to derive the thermal contribution. {\it Right:} Turbulent Doppler parameter distribution at $z=3$ as predicted by our model (histogram), also reporting 1-$\sigma$ dispersion. The red arrows show the observational upper limits.}
  \label{comparison}
\end{figure*} 

A direct comparison of our results with the actual level of turbulence in the IGM is not possible since this observable has not been convincingly derived yet. In the meantime, it is possible to perform at least a sanity check for our model as follows.  

In order to study the properties of the IGM, a number of high-quality QSO absorption line experiments has been carried on during the last decade allowing the build-up of a large sample of observed systems for which both $N_{\rm HI}$ and the Doppler parameter $b$ have been fitted (see an almost complete list of references in \citealp{Meiksin09}). The observed HI column densities cover the range $10^{12-22}$~cm$^{-2}$ and follow an almost perfect power-law distribution $\de N/\de N_{\rm HI} \propto N_{\rm HI}^{-\beta}$ with $\beta = 1.5 - 1.7$ . The Doppler parameters of these systems range over about $10<b<100$~km~s$^{-1}$, with the vast majority being concentrated between $15-60$~km/s. The observed broadening could be either real, i.e. caused by physical mechanisms (e.g. thermal broadening, turbulence and peculiar motions), or artificial, due to the blending of close systems.

Many authors have suggested the use of the lower-cutoff in the Doppler parameter distribution vs. $N_{\rm HI}$ as a reliable proxy for the gas temperature at a given redshift. Similar approaches has been used by \citet{Schaye00b} to report evolution in the inferred IGM gas temperature over the range $2.0 < z < 4.5$ and by \citet{Ricotti00} to extract the \emph{equation of state} of the IGM from the Doppler parameter distribution of the low-density Ly$\alpha$ forest. To make progress we follow this procedure. 

In Fig.~\ref{comparison} we show the absorber properties derived in a sample of Ly$\alpha$ forests from 22 high-resolution VLT/UVES Large Program QSO spectra \citep{Dodorico08,Saitta08}, selected at redshift $z\sim 3$. To avoid the contamination of absorbers affected by the UV background flux from the QSO, we exclude from our sample absorbers closer than $3000$~km/s to the QSO and we also do not consider in our analysis Doppler parameters $>100$~km~s$^{-1}$ since they are very likely due to errors in the fitting procedure. In the same graph we report our estimated thermal Doppler parameter ($b_{\rm th}$) as a function of $N_{\rm HI}$, obtained from a linear fit of the {\it minimum} $b$-values, defined by dividing the $N_{\rm HI}$ range in 50 bins and selecting the smaller b-value in each bin.

The sanity check then consists in the comparison of our predicted mean $b_t$ distribution with the analogous average residual Doppler parameters from the data after the thermal contribution has been subtracted. Given the reasons for contamination listed above what we show in Fig.~\ref{comparison} cannot be intended as a direct theory-data comparison as the observed broadening still contains a contribution from IGM peculiar motions, which are not directly related to turbulent motions. Hence, the data points must be considered as strict upper-limits which our theoretical Doppler parameters should not exceed. Quite encouragingly, this is indeed the case: our model is consistent with the experimental upper limits in the entire HI column density range. The uncertainties in the exact observational determination of $b_t$ does not allow a more meaningful comparison at this stage. This will become possible in the future if Rauch-type experiments using (lensed) quasar pairs will be carried out. The increased availability of such systems, enabled by the SDSS, has been already shown by \citet{Kirkman08}, who have discovered 130 close pairs of QSOs in the SDSS DR5.

As an alternative, the presence of the metal lines in the absorbers could also be used as a possible way to obtain an independent measurement of the IGM turbulent level. The measurements of the Doppler parameters from absorption features corresponding to two or more elements (typically, CIV and SIV) assumed to be co-spatial, allows the separation between thermal and kinematic broadening contributions. As already mentioned in the previous Section, this technique could disentangle the broadening due to the kinematic from the thermal one which is instead mass dependent. To achieve this result high resolution spectroscopy is required in order to properly fit the profile of the absorbers along the QSO line of sight and to determine the redshift of the lines with sufficient precision to make sure that the ions share the same absorber environment. We plan to combine these type of observations with our model in a forthcoming work. 

\section{SUMMARY AND IMPLICATIONS}

Turbulence is an important physical process in essentially all astrophysical environments, from planet formation and atmospheres, to (proto-)stars and galaxies, but particularly 
on the scale relevant to cosmological phenomena it has received a relatively meager attention. This is partly understood from the difficulty to collect reliable data as observations aimed at the study of these aspects are very challenging. Even more surprisingly though, the well ascertained fact that the IGM is polluted on large scales by heavy elements, carried by powerful winds powered by galaxies, should raise strong suspicions that turbulence must pervade the IGM, albeit at levels yet difficult to understand.

The present paper aims at making one of the first steps to clarify the amount of turbulent energy deposited in the IGM and then dissipated through a cascade to smaller scales by galactic winds and to characterize its properties. To this aim we have developed a simple but physically accurate model based on standard \lcdm hierarchical structure formation model to identify halos hosting galaxies progenitor of a small cluster/group of total mass $M=10^{13} M_\odot$ at $z=0$. In addition, the required star formation history in all their progenitors via a merger-tree technique (carefully calibrated to reproduce the Milky Way) is also followed. As these galaxies progress along their history, their collective supernova explosion vent gas, heavy elements and, most importantly for the present study, energy into the surrounding IGM. The evolution of these hot intergalactic bubbles is followed by including all the relevant physics, enabling us to determine their thermodynamic properties at best. We couple then this general galaxy-IGM interplay scenario (which can be used and adapted to a number of ancillary applications as e.g. metal enrichment and SZ effect) to a treatment of turbulence evolution using a powerful approach based on the spectral transfer function developed by \citet{Norman96}. In brief, the major advantage of this method is a relatively straightforward derivation of the turbulent IGM turbulent pressure evolution (see Eq.~\ref{pturb}). 

The main results can be summarized as follows. At $z\approx 3$ the majority of the bubbles around galaxies are old (ages $> 1$~Gyr), i.e. they contain metals expelled by their progenitors at earlier times, and have a size distribution at that redshift in the range 10-100 (physical) kpc. The velocities reach up to $100$~km~s$^{-1}$ for larger galaxies but a considerable fraction of the bubbles have already become subsonic and the shock decayed into a sound wave; temperatures in the rarefied bubble cavities are in the range $5.4 < \log (T/K) <6.5$. The bubble volume filling factor increases with time reaching about 10\% at $z < 2$. The energy deposited by these expanding shocks in the IGM is predominantly in kinetic form (mean energy density of 1 $\mu$eV cm$^{-3}$, about 2-3$\times$ the thermal one), which is rapidly converted in disordered motions by instabilities, finally resulting in a fully developed turbulent spectrum. The corresponding {\it mean} turbulent Doppler parameter, $b_t$, peaks at $z\approx 1$ at about 1.5~km~s$^{-1}$ with maximum $b_t = 25$~km~s$^{-1}$. More informative though is the $b_t$ distribution associated with individual bubbles and the relations with the galaxy stellar mass and bubble shell column density and radius (shown in Fig. 5). The shape of the $b_t$ distribution does not significantly evolve with redshift but undergoes a continuous shift towards lower $b_t$ values with time, as a result of bubble aging. We find also a clear trend of decreasing $b_t$ with $N_{\rm HI}$ and a more complex dependence on $R_s$ resulting from the age spread of the bubbles. We have finally tried to compare our results with available data on the Doppler parameter deduced from QSO absorption line studies of the Ly$\alpha$ forest, but we are facing the difficulty of accurately removing the contamination due to the thermal component and the more troublesome peculiar contribution. Alternatively one can extend our study to make predictions for quasar pairs systems and/or using observations involving different metal species. We are already working along these directions, which in general require additional information of the position of galaxies and therefore the implementation of our turbulent model into a full-scale hydrodynamic simulations.       

If turbulent is indeed present in the intergalactic medium, as our study seems to suggest, it might have relevant implications for a number of IGM studies. First of all the long-standing problem of a discrepancy between the observed Doppler parameter observed and those deduced from numerical simulations of the Ly$\alpha$ forest might be alleviated by a proper inclusion of turbulence broadening. \citet{Oppenheimer09} convincingly demonstrated that only a model where explicitly add sub-resolution turbulence, and the required turbulence increases with O VI absorber strength, is able to match all the observations, giving an hint of this correlation between turbulence and enrichment that we are planning to explore in a forthcoming paper. Secondly, the turbulent energy injection might affect the matter power-spectrum determination from experiments using the Ly$\alpha$ forest as a probe, particularly on the smallest scales \citep{McDonald06}. 
Third, if turbulence is largely present in the IGM it might result in a detectable signature in the scintillation of distant quasars \citep{Ferrara01,Lazio04}. The IGM hosts the necessary conditions for scattering, is largely ionized and permeated by shocks from large scale formation or galactic feedback. Moreover, the long path lengths through the IGM can compensate for the lower density with respect to the ISM. \citet{Lazio08} have considered intergalactic scattering, expanding on previous treatments from intra-cluster media of \citet{Hall79}, to scintillating AGNs, their findings are broadly consistent with the scenario of a highly turbulent intergalactic medium. This possibility was explored so far only as an {\it Ansatz}, to which our study now provides a more solid support. Fourth, turbulence might be carefully modeled in order to interpret the results for experiments aiming at directly measuring the change of the expansion rate of the Universe with time and the variability of fundamental constants \citep{Corasaniti07,Liske08}, as for example the proposed CODEX-ESPRESSO experiment \citep{Cristiani07}. Last but not least, the relatively mild level of turbulence we find at $z\approx 3$ is consistent so far with the upper limits coming from observations. Moreover, it indicates that the IGM at those epochs had already dissipated a large fraction of the kinetic energy  deposited at early times ($z>5$) by the galaxies which initiated the metal enrichment and, possibly, the reionization process. All this well agrees with the idea that the metals we detect at moderate redshifts were mostly produced during a "pre-enrichment" phase of the IGM, by a population of small galaxies ancestors of the ones we study routinely in detail, the  Lyman Break Galaxies. 

\section*{Acknowledgments}
We thank V. D'Odorico for useful suggestions and for supplying the Ly$\alpha$ forest data we used in our analysis. Stimulating conversations with P. Petitjean, J. Niemeyer and R. Valdarnini are warmly acknowledged. CE would like to thank Scuola Normale Superiore in Pisa for the warm hospitality during the preparation of this work.

\bibliographystyle{mn2cls/mn2e}
\bibliography{mn2cls/mn-jour,EF10}

\bsp

\label{lastpage}

\end{document}